# Nd-doped polarization maintaining all-fiber laser with dissipative soliton resonance mode-locking at 905 nm

Aram A. Mkrtchyan, Yuriy G. Gladush, Mikhail A. Melkumov, Aleksandr M. Khegai,
Kirill A. Sitnik, Pavlos G. Lagoudakis, Albert G. Nasibulin

*Abstract*—**Moving the fiber laser emission to the region below one micron may provide a cheaper, more compact and robust alternatives to the existing solid state lasers. Here, for the first time we report a neodymium mode-locked fiber laser emitting at 905 nm in the all-fiber polarization maintaining configuration. We obtain a self-starting pulse generation in nonlinear amplifying loop mirror (NALM) cavity configuration. To suppress a dominant emission at 1064 nm corresponding to a 4-level laser scheme, we use an active fiber – 920/1064 division multiplexer – active fiber sandwich-like sequence in the NALM loop. A rectangular shape dissipative soliton had nJ energy, 30 pm spectral width and 80 ÷ 430 ps width linearly depending on the pump power. Excellent agreement with numerical simulation allowed us to recover pulse shape and width for the pulses out of autocorrelation window.**

I. INTRODUCTION

ULTRAFAST fiber lasers attracted much attention in recent years because of their compactness, low cost, robustness and simplicity of operation, and considered as an ideal light source for two-photon microscopy (TPM), optical coherent tomography, laser spectroscopy, precision metrology, LIDAR, quantum communications and also for generation of blue light through frequency doubling [1]–[4]. However, most of the fiber systems work in the spectral region between 1 and 2 μm [5]–[13]. Wavelength below 1 μm remains domain of the solid-state lasers, with the domination of titanium-sapphire (Ti:Sa) lasers with excellent ultrashort pulse emission tunable in 690–1050 nm wavelength range. The TPM at submicron wavelengths has undergone tremendous development since the first demonstration in 1990 [14] and has been applied in cell biology and neurosciences. Traditionally, the TPM employs Ti:Sa lasers as a light source for practical clinical applications [15]. However, bulky size and expensive cost of the Ti:Sa lasers limit their wide spreading beyond laboratories. The TPM at the 900 nm spectral window enables imaging living cells, two-photon excitation of a variety of green fluorescent proteins, spurring the development of integrated all-fiber pumping sources at ∼900 nm.

Recently, few techniques were utilized to generate ultrashort pulses below 1 μm: femtosecond Yb-doped fiber laser amplification with optical parametric oscillator [16], [17], nonlinear frequency shift in a high-nonlinearity fiber [18], [19], frequency doubling of the 1.8-1.9 μm ultrafast laser emission [20], fiber-optic Cherenkov radiation technique inside a short piece of photonic crystal fiber [21] and direct generation using the neodymium-doped fiber (Nd-fiber) quasi-three-level $^4F_{3/2}$ - $^4I_{9/2}$ transition [22]–[31]. The last one is preferred due to its robustness and simplicity. However, stimulated emission efficiency for the Nd-fiber in 900 nm range is an order of magnitude lower than the efficiency for four-level transition $^4F_{3/2}$ - $^4I_{11/2}$ at 1064 nm. Few approaches were proposed to suppress emission at 1064 nm. First, the Nd-fiber with a length as short as 2.5 cm was used to decrease the difference of the emission gain at this wavelength range [32]. Integrating short Nd-fiber into a linear laser scheme with a single-frequency distributed Bragg reflector enhanced the emission at ∼ 900 nm range. In this case only continuous-wave operation with 1.8% efficiency was demonstrated. Second, grating pair with adjacent collimators with limited aperture size was implemented, causing strong spectrum filtering at 1060 nm and forced the laser to work at ∼900 nm [24], [26]. Another approach to suppress the strong laser oscillation at ∼1064 nm is to insert into the cavity a short-pass dichroic mirror with high transmission at ∼900 nm and high reflectivity at ∼1064 nm [22], [23]. *Rusu et al.* have utilized original semiconducting

Acknowledgments; The reported study was funded by RFBR according to the research project № 20-32-90233. *(Corresponding author: Aram A. Mkrtchyan.)*

A. A. Mkrtchyan, Y. G. Gladush and K. A. Sitnik are with the Skolkovo Institute of Science and Technology, Nobel 3, Moscow 121205, Russia (e-mail: Aram.Mkrtchyan@skoltech.ru; Y.Gladush@skoltech.ru; K.Sitnik@skoltech.ru).

M. A. Melkumov, A. M. Khegai are with the Prokhorov General Physics Institute of the Russian Academy of Sciences, Dianov Fiber Optics Research Center, Moscow, 117942, Russia (e-mail: melkoumov@fo.gpi.ru; hegayam@gmail.com).

P. G. Lagoudakis is with the Skolkovo Institute of Science and Technology, Nobel 3, Moscow 121205 and Department of Physics and Astronomy, University of Southampton, Southampton SO17 1BJ, United Kingdom (e-mail: P.Lagoudakis@skoltech.ru).

A. G. Nasibulin is with the Skolkovo Institute of Science and Technology, Nobel 3, Moscow 121205, Russia and Aalto University, Department of Chemistry and Materials Science, Aalto 00076, Finland (e-mail: A.Nasibulin@skoltech.ru).

saturable absorber mirror (SESAM) including 20 pairs of AlAs – $Al_{0.2}Ga_{0.8}As$ quarter-wave layers and forming a distributed Bragg reflector with 100-nm reflection bandwidth (860–960 nm) to suppress emission at 1064 nm [27]. Finally, neodymium doped (Nd-doped) double-clad fiber with a W-type core refractive index profile was implemented as a gain medium. It creates an effective cutoff wavelength of the $LP_{01}$ mode at longer wavelengths that led to this mode suffer from high losses distributed along the fiber, resulting in sufficient decrease of the gain at 1060 nm [28], [30], [33]. The works mentioned above have successfully demonstrated ultrafast operation based on the nonlinear polarization evolution, nonlinear amplifying loop mirror (NALM) or SESAM. However, all of them have contained non-fiber free-space optical components, making them sensitive to the mechanical perturbations and limiting their implementation in practical applications.

Here, for the first time we demonstrate all-fiber polarization maintaining mode-locked rectangular shape pulse laser operating at 905 nm wavelength in the NALM scheme. Single clad Nd-doped fiber was utilized as an active medium in the laser resonator. To filter out longer wavelengths, we utilized 920/1064 nm wavelength division multiplexer (WDM) sandwiched between two pieces of active fibers, leading to efficient suppression of parasitic emission at 1064 nm. Thus, the laser operates at the desired wavelength with more than 50 dB dominance over the parasitic one. The laser demonstrated self-starting stable nJ energy pulse generation with $80 \div 430$ ps duration linearly depending on the pump power and minimum 30 pm spectral width. In the frame of this work we carried out numerical simulation and showed perfect correspondence of the theoretical and experimental results at large normal dispersion corresponding to the dissipative soliton resonance regime.

## II. EXPERIMENTAL SETUP AND NUMERICAL SIMULATIONS

All-PM fiber laser schematically shown in Fig. 1a. The mode-locking was obtained with figure-eight laser scheme with nonlinear amplifying loop mirror mechanism. The laser consisted of main and NALM loops. The main loop included isolator (ISO) with blocked fast axis, determining lasing direction and polarization, and an output coupler extracting 10% of emission power from the laser. The laser emission propagated from the main to the NALM loop via X-type coupler dividing a light into two counterpropagating beams in 49.7/50.3 ratio at 905 nm. The NALM loop contained two pieces of active fibers and 920 nm/1064 nm wavelength division multiplexer (WDM) sandwiched between them followed by 27 m of passive PM-850 fiber. Two pieces of $L_{Nd}$ = 1.3 m long active fibers were from single clad Nd-doped fiber: CorActive ND 103-PM, with ≈ 40 dB/m core absorption at 808 nm and core pumped with laser diode at 808 nm (LU0808M250 Lumics) through 800 nm/920 nm WDM. The splitting ratio of X-type coupler was enough for self-starting of pulse generation in the laser. All components were PM ensuring the linearly polarized stable operation (for output power stability see Fig. 1a inset). The laser cavity had net normal dispersion as large as 2.4 ps$^2$ with total cavity length $L_{tot}$ = 34.2 m ($L_{main}$ = 4.2 m main loop and $L_{NALM}$ = 30 m NALM loop). All fibers estimated to have 70 fs$^2$/mm dispersion at 905 nm, which was obtained with best correspondence of simulated and experimentally measured pulse spectrum and profile.

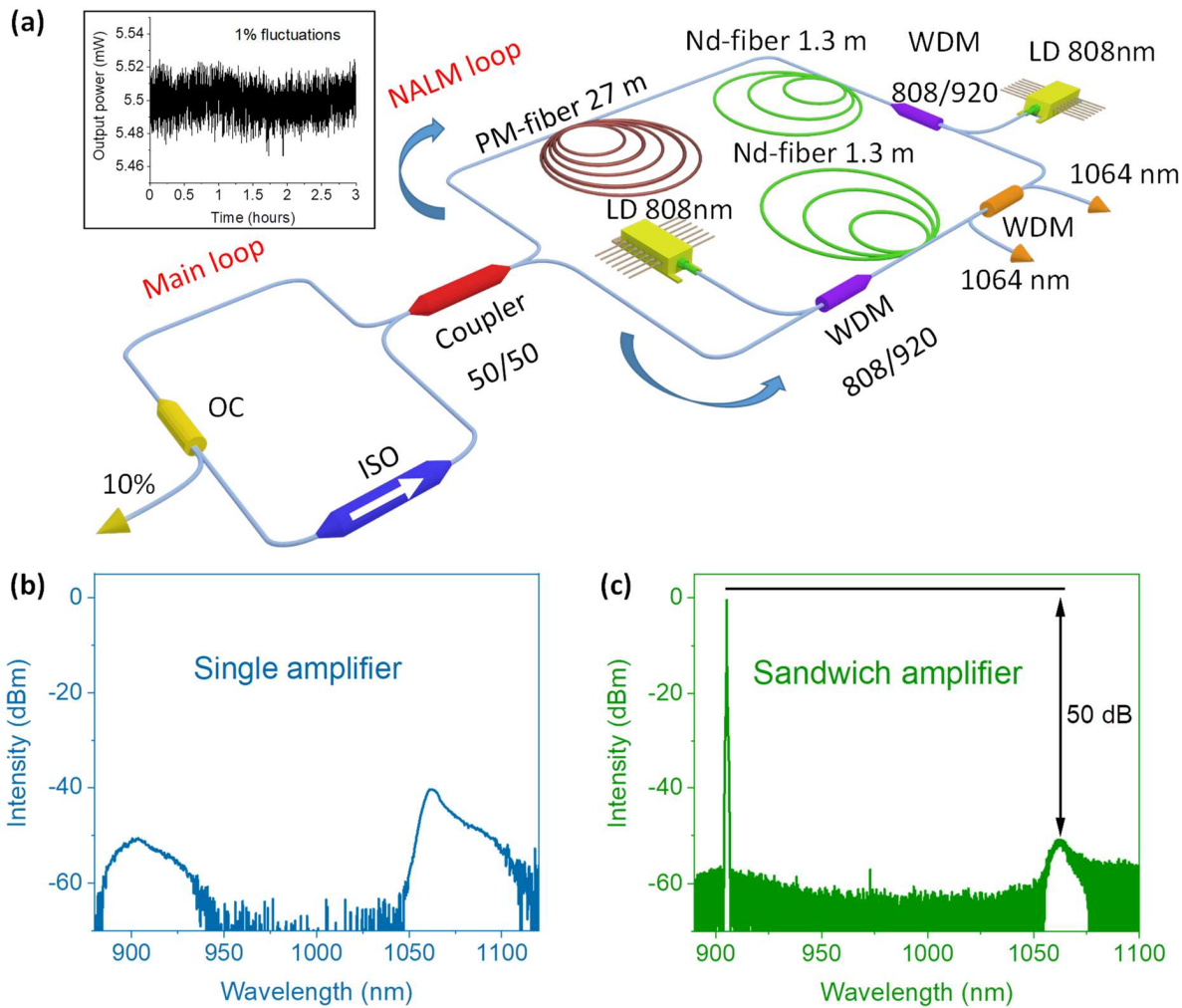

Fig. 1. a) Nd-doped polarization maintaining all-fiber dissipative soliton mode-locked laser: output coupler (OC), isolator (ISO), neodymium doped fiber (Nd-fiber), wavelength division multiplexer (WDM), laser diode (LD). Inset demonstrates an output power stability during 3 hours with around 1% signal fluctuation. b) Amplified spontaneous emission of the laser with single amplifier. c) Lasing at 905 nm with 50 dB dominance over 1064 nm.

To filter out the emission at 1064-band we introduced an active fiber-spectral filter-active fiber (ASFA) sequence into the NALM loop. In case of single active fiber with equivalent amplification to the ASFA and single spectral filter, 1064-band cannot be filtered out completely because light propagates in both directions in the loop. The light propagation in the direction opposite to filter returns to the active fiber through the main loop giving rise to the amplified spontaneous emission (ASE) at 1064 nm (Fig. 1 b). In contrast, in the ASFA sandwich case 900/1064 WDM, located between two active fibers, filters out 1064-band, leaving 900-band to seed laser generation in the opposite active fiber. The 900-band had a maximum at 905 nm leading the laser to operate at this wavelength with 50 dB dominance over 1064-band (see Fig. 1c).

The obtained pulsed laser characteristics were measured with the following equipment: Femtochrome FR-103WS autocorrelator, 10 pm resolution spectrum analyzer YOKOGAWA AQ6373B, the oscilloscope MDO3000 Tektronix and built-in radiofrequency (RF) analyzer having 30 Hz resolution bandwidth and 5 GHz photodetector Thorlabs DET08CFC, slim photodiode power sensor S132C.

We also carried out numerical simulations of the figure-eight pulse fiber laser shown in the Fig. 1a. Signal propagation through the laser cavity were described by the Ginzburg-Landau equation for complex amplitude [34], [35]:

$$\frac{\partial A}{\partial z} - i\frac{\partial \beta_{2g}}{2}\frac{\partial^2 A}{\partial t^2} - i\gamma_g |A|^2 A = \frac{gA}{2} + \frac{\partial \beta_{2f}}{2}\frac{\partial^2 A}{\partial t^2}. \quad (1)$$

Here $A(z,t)$ is the electric field in slowly varying amplitude, $z$ is the coordinate along laser cavity and $t$ is time in an accompanying coordinate, $\beta_{2g}$ corresponds to the second order group velocity dispersion in the PM-850 fiber near 905 nm wavelength, $\gamma$ is the Kerr nonlinearity. Here $\beta_{2f} = g/\Omega^2$ corresponds to the parabolic spectrum gain with $\Omega$ width and $g$ is the saturable gain:

$$g(z,t) = g(z) = g_0 \left(1 + \frac{\int_0^{\tau_{win}} |A(z,t)|^2 dt}{E_g}\right)^{-1} \quad (2)$$

Here $g_0$ is the small signal gain, $\tau_{win}$ is simulation window in the time domain and $E_g = T_{RT} * P_{sat}$ gain saturation energy, where $T_{RT}$ is the roundtrip time through the fiber laser and $P_{sat}$ is the gain saturation power. The nonlinear Schrödinger equation was used for propagation through the passive fiber:

$$\frac{\partial A}{\partial z} - i\frac{\partial \beta_2}{2}\frac{\partial^2 A}{\partial t^2} - i\gamma |A|^2 A = 0. \quad (3)$$

Light propagation in two directions inside the NALM loop was simulated separately also using Ginzburg-Landau equation and the interference of these signals determined the transmission coefficient of the NALM loop. The group velocity dispersion and Kerr nonlinearity parameters were estimated to be the same for both active and passive fibers.

Equations (1-3) have been simulated by the split-step Fourier method in the simulation window τwin discretized with 2$^{13}$

TABLE I
NUMERICAL SIMULATION PARAMETERS

| Parameter | Value | Units | Description |
|---|---|---|---|
| $\beta_2$, $\beta_{2g}$ | 70 | fs$^2$/mm | 2$^{nd}$ order group velocity dispersion[a] |
| $\gamma$, $\gamma_g$ | 0.01 | W$^{-1}$m$^{-1}$ | Kerr nonlinearity[a] |
| $\Omega$ | 49 | nm | parabolic spectrum gain width |
| $\tau_{win}$ | 1 | ns | simulation window |
| $P_{sat}$ | 20 | mW | gain saturation power[a] |
| $\alpha$ | 6.4 | dB | total losses |
| $g_0$ | 3.5÷11.5 | dB | small signal gain |
| $L_{NALM}$ | 30 | m | NALM loop length |
| $L_{main}$ | 4.2 | m | main loop length |
| 2 x $L_{Nd}$ | 2.6 | m | total Nd-doped fiber length |

[a]Estimated parameters.

points. Total losses of the laser were about α ≈ 6.4 dB coming from the laser components losses and output coupler. A low-amplitude Gaussian spectral noise was defined as the initial condition. The values of the parameters were selected to ensure best correspondence between experimental results and simulation and are shown in Table 1.

The output was examined after 2000 roundtrips, which ensured stationary pulse generation. The results of the numerical simulation will be described in the next section.

III. RESULTS AND DISCUSSIONS

This all-fiber laser has generated dissipative soliton pulses with a rectangular shape in the time domain and with a narrow spectrum at full width at half maximum (FWHM) on the wider pedestal, which is common for the dissipative solitons resonance regime [36], [37]. The NALM laser demonstrated self-starting and mode-locking at 370 mW of total pump power and showed stable operation at lower pump power decreased down to 210 mW. The average output power and power respectively reaching 7.5 mW and 1.24 nJ at maximum available 562 mW pump power while keeping operation in the single pulse regime (see Figure 2a). pulse energy of the laser had linear dependence on the pump The optical spectra of the pulses (see Fig. 2b) had narrow linewidth decreasing with the increase of the pump power from 0.14 nm to as narrow as 30 pm at FWHM measured with 10 pm resolution spectrometer (inset in Fig. 2b). The pedestal in the spectra is observed at -20 dB level from the maximum. It is also shown in numerically simulated spectra of the pulse with the minimal energy of 0.25 nJ (dashed line in Fig. 2c), which shows excellent correspondence with the experimental data (solid line in Fig. 2c).

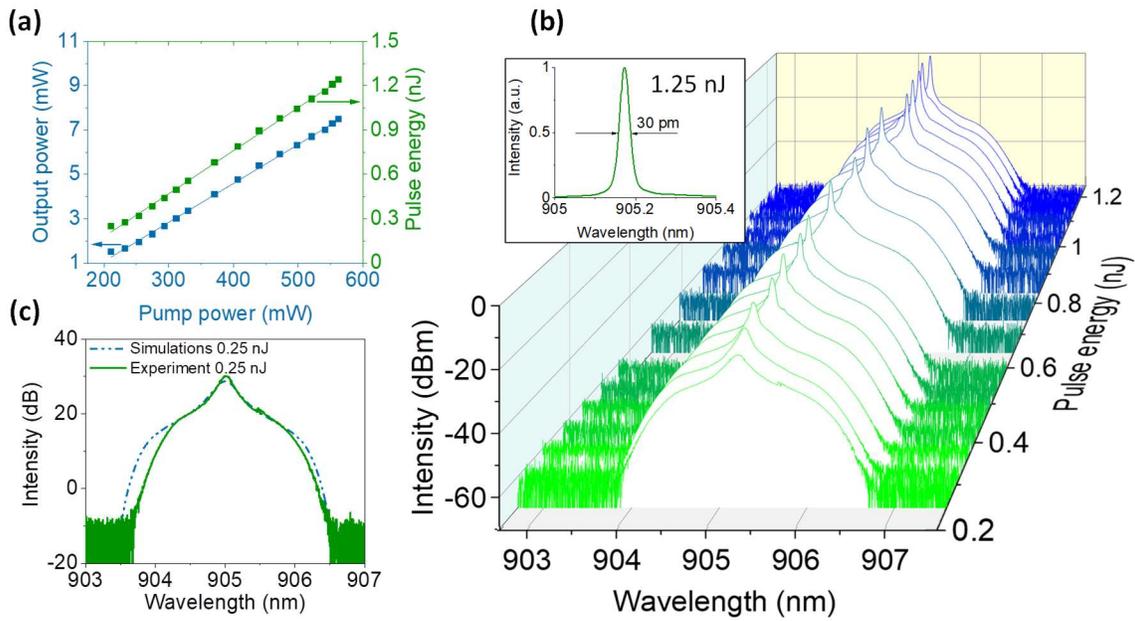

Fig. 2. a) Average output power and generated pulse energy dependence on the pump power. b) Spectra of generated pulses at different pulse energy; inset shows 30 pm spectrum width of the pulse at 562 mW pump power. c) Comparison of the experimental and numerically simulated spectra of 0.25 nJ energy pulse.

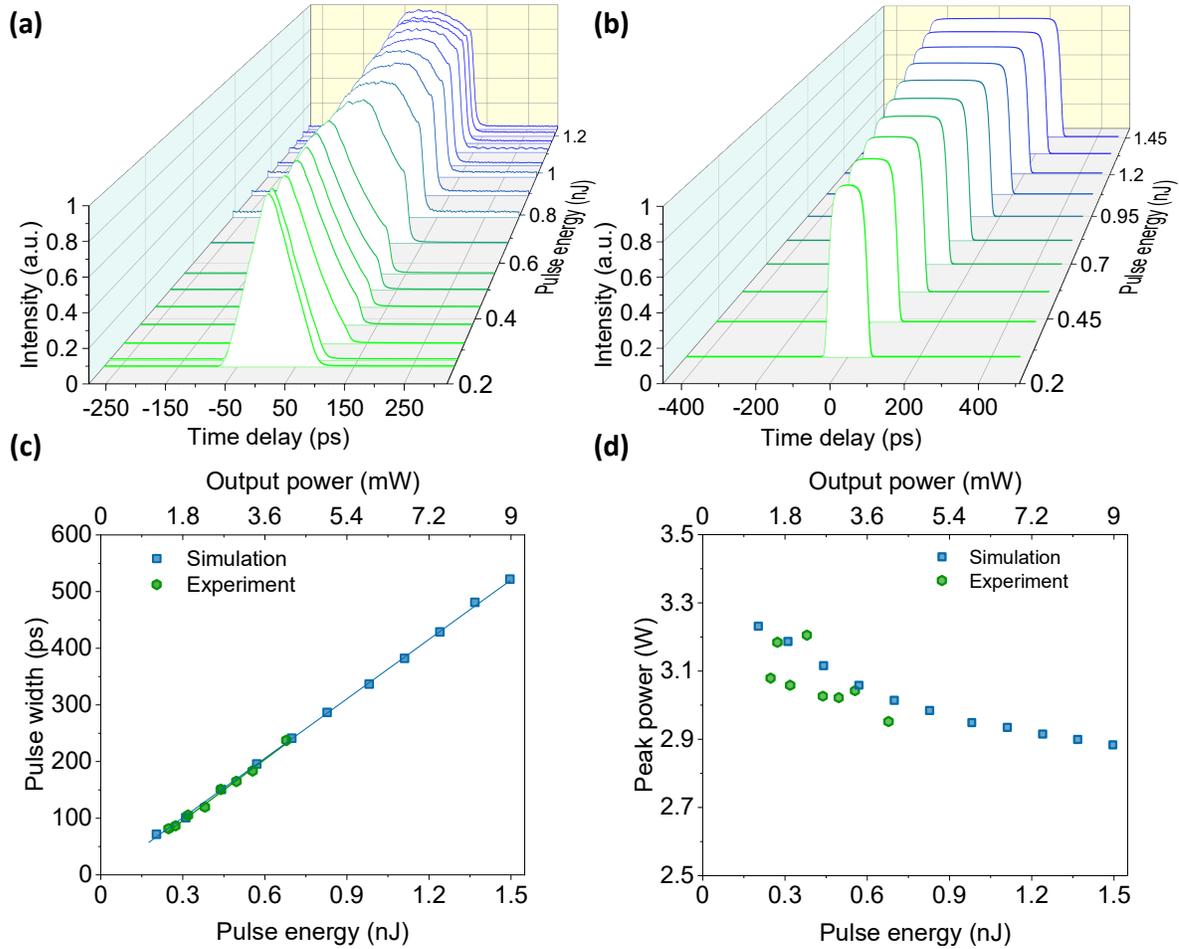

Fig. 3. a) Experimental autocorrelation traces. Pulse with energies exceeding 0.7 pJ are outside of autocorrelator scanning window and their autocorrelation function is distorted. b) Simulated pulse shapes of dissipative solitons. Measured and calculated c) pulse width and d) pulse peak power versus the pulse energy.

The autocorrelation functions of generated pulses (see Fig. 3a) had clear triangular shape corresponding to the rectangular shape of the pulses in time domain. In the case of rectangular pulses, the autocorrelation trace full width at half maximum (FWHM) equal to the pulse FWHM. The measured pulse durations were in the ps range linearly varying from 80 ps to the 263 ps width in 211 ÷ 407 mW pump power range. The operating window of the autocorrelator was 250 ps limiting the measurement of the wider autocorrelation function. To estimate the pulse durations for higher pump power we numerically simulated the pulses for a wide pulse energy range. The simulated pulse envelopes with rectangular shapes are shown in Fig. 3b. Here, we calculated the pulse widths depending on the pulse energy. Experimental and numerical pulse widths were showed excellent agreement in the region where pulse duration can be measured experimentally (see Fig. 3c). For higher pulse energies, the pulse duration were obtained from numerical simulations and showed that at the maximum available 562 mW pump power pulse width was nearly 430 ps. From experimental data and numerical simulations, we calculated the peak power of pulses to be nearly 3 W that slightly decreases when increasing the pump power (see Fig. 3d). Clamping of the peak power is usually observed for dissipative soliton pulses [38] and mostly defined by the NALM loop length, so that the phase shift between two counter propagating beams inside the NALM loop is optimal. The GVD had very little impact on the temporal pulse shape affecting mainly the width of the pulse spectrum pedestal. Longer NALM loop length leads to smaller peak power, and thus wider pulse width at the same pump power [31].

The pulse train had 165 ns repetition time (see Figure 4a) corresponding to repetition rate of 6 MHz (see Figure 4b) and 34.2 m laser cavity length. 65 dB signal to noise ratio of the radiofrequency spectrum proves the stability of the generated pulses in the laser.

The rectangular shape in time domain, the narrow optical spectrum with the pedestal and the nearly fixed peak power perfectly correlate with the dissipative soliton theory for large net dispersion range being 2.4 ps$^2$ in our all fiber laser [36].

## IV. Conclusion

In conclusion, we demonstrated all PM mode-locked Nd-doped fiber laser operating at the wavelength of 905 nm. Rectangular shape dissipative solitons were obtained with nonlinear amplifying loop mirror mechanism in a large 2.4 ps$^2$ normal dispersion regime. We utilized spectral filtering mechanism with the wavelength division multiplexer sandwiched between two neodymium doped single clad active fibers, ensuring suppression of parasitic emission at 1064. We obtained output pulse generation at 905 nm wavelength with a 50 dB dominance over ASE level. The laser demonstrated self-starting with a stable nJ energy pulse generation, 80 ÷ 430 ps duration and narrow 30 pm spectral width. Also, obtained dissipative solitons were analyzed numerically and showed excellent agreement with the experimental results. From the numerical simulation we estimated the shape of the pulses and its width and found their peak power being nearly 3 W and slightly decreasing with increase of the pump power. Such narrow linewidth pulses with hundreds of ps duration could be an effective source for two-photon microscopy enabling the imaging of living cells, efficient frequency doubling and for excitation of single photon sources.

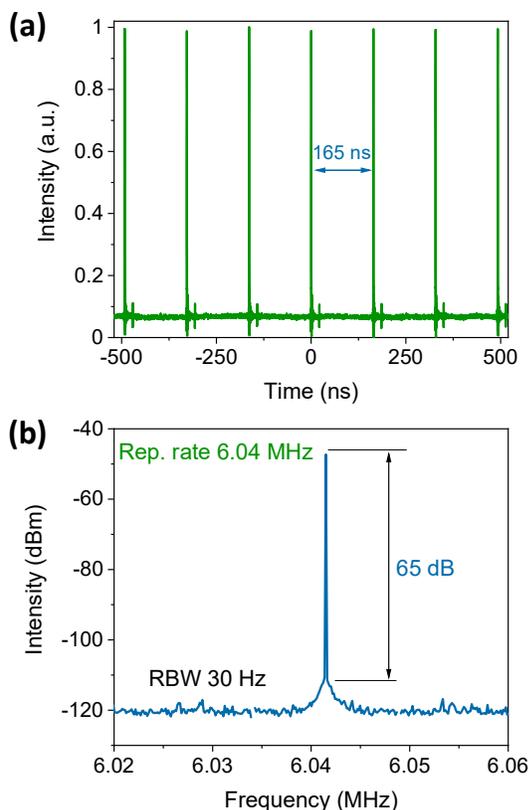

Fig. 4. a) Pulse train with 165 ns delay between pulses, b) Radiofrequency spectrum of the pulses with 6.04 MHz fundamental repetition rate and signal-to-noise ratio exceeding 65 dB measured with 30 Hz resolution bandwidth.

**Aram A. Mkrtchyan** was born in Yerevan, Armenia on Nov. 3, 1994. He received the B.S and M.S. degrees in applied mathematics and physics from Moscow Institute of Physics and Technology, Moscow, Russia and Skolkovo Institute of Science and Technology, Moscow, Russia. He is currently pursuing the Ph.D. degree in laser physics at Skolkovo Institute of Science and Technology. His research interest includes the development of ultrafast all-fiber lasers, at wide spectral ranges including submicron wavelengths. He is a winner of RFBR grant for Ph.D. student, where is working on development of optical devices, with controllable nonlinearity based on carbon nanomaterials for applications in fiber lasers: ultrafast generation with controllable pulses regimes, second harmonic generation, four-wave mixing.

**Yuriy G. Gladush** was born in Moscow, Russia, on Aug. 16, 1983. He received the Ph.D. degree in theoretical investigation of nonlinear wave phenomena in Bose-Einstein condensates and optics from Institute for Spectroscopy Russian Academy of Sciences, Troitsk, Moscow.
Gladush is an Assistant Professor in Laboratory of Nanomaterials of Skolkovo Institute of Science and Technology. His main research area is optics and laser physics. In Nanomaterials lab he is responsible for projects related to optical applications of carbon nanotubes and other nanomaterials including fiber laser ultrashort pulse generation and SWCNT bolometer development. Prior to Skoltech prof. Gladush was working in Institute of spectroscopy where his research was dedicated to resonance energy transfer in organic/inorganic semiconductor hybrid structures.

**Mikhail A. Melkumov** was born in Moscow, Russia, on Nov. 21, 1978. He received the Graduate degree from the Physics Department, Moscow State University, Moscow, Russia, in 2001, and the Ph.D. degree from the Fiber Optics Research Center, Russian Academy of Sciences, Moscow, Russia, in 2006. Currently, he is the Head of the Fiber Lasers and Amplifiers Laboratory, Prokhorov General Physics Institute of the Russian Academy of Sciences, Dianov Fiber Optics Research Center. His research interests include Raman and rear-earth-doped fiber lasers, bidoped fiber lasers and amplifiers, and spectroscopy of active centers in silica-based glasses and fibers.

**Aleksandr M. Khegai** was born in Sovietabad, Andijan region, Russia, on Apr. 23, 1991. He received the B.S. and M.S. degrees in physics from Volgograd State Technical University, Volgograd, Russia, in 2014. He is currently working toward the Ph.D. degree in laser physics in Prokhorov General Physics Institute of the Russian Academy of Sciences, Dianov Fiber Optics Research Center. His research interests include spectroscopy of the rare-earth- and bismuth-doped active fibers, the development of continuous wave and ultrafast fiber lasers and amplifiers.

**Kirill A. Sitnik** was born in Novokuznetsk, Kemerovo region on 24th January 1991. He received B.S and M.S. degrees in 2012 and 2014 respectively in National Research Tomsk State University. Later he had worked in Institute of High-Current Electronics of Siberian Branch of Russian Academy of sciences as engineer in Laboratory of Gas Lasers. Currently he is 3rd year PhD student in Laboratory of Hybrid Photonics in Skoltech. Main direction of his research is quantized vortices in polariton condensate.

**Pavlos G. Lagoudakis** graduate of the University of Athens, Greece. Pavlos received his PhD degree in Physics from the University of Southampton, UK in 2003 and conducted his postdoctoral research on optoelectronic properties of organic semiconductors at the Ludwig Maximilians University of Munich, Germany. In 2006, he returned to Southampton as Lecturer at the department of Physics and Astronomy, where he combined his expertise in inorganic and organic semiconductors and set up a new experimental activity on Hybrid Photonics.

In 2008, Lagoudakis was appointed to a personal chair at the University of Southampton. From 2011 to 2014, Pavlos chaired the University's Nanoscience Research Strategy Group, an interdisciplinary research group of ~100 academics across physics, chemistry, maths, engineering, biology and medicine. From 2013 to 2019, Pavlos was the Director for Research at the department of Physics and Astronomy at the University of Southampton.

At Skoltech, prof. Lagoudakis designed, setup and now consults the Hybrid Photonics Group with a focus on hybrid LEDs, PVs and spinoptronics.

**Albert G. Nasibulin** was born in Novokuznetsk, Kemerovo Region, Russia on March 23, 1972. He got his PhD in Physical Chemistry (1996) at Kemerovo State University (Russia) and Doctor of Science (Habilitation, 2011) at Saint-Petersburg Technical State University (Russia). He is a Professor at Skolkovo Institute of Science and Technology and an Adjunct Professor at the Department of Chemistry and Materials Science of Aalto University School of Chemical Engineering. He held a post of the Academy Research Fellow in Academy of Finland from 2006 to 2011. Since 2018 he is a Professor of the Russian Academy of Sciences.

Prof. Nasibulin has specialized in the aerosol synthesis of nanomaterials (nanoparticles, carbon nanotubes and tetrapods), investigations of their growth mechanisms and their applications. He has a successful background in an academic research with about 300 peer-reviewed scientific publications and 40 patents. He is a co-founder of three companies: Canatu Ltd. (spin-off from Helsinki University of Technology, Finland) and CryptoChemistry and Novaprint (spin-offs from Skolkovo Institute of Science and Technology, Moscow, Russia).